\documentclass{article}

\usepackage{microtype}
\usepackage{graphicx}
\usepackage{subfigure}
\usepackage{booktabs} 

\usepackage{hyperref}

\usepackage{tcolorbox}
\tcbuselibrary{listings}

\usepackage[accepted]{icml2023}

\usepackage{amsmath}
\usepackage{amssymb}
\usepackage{mathtools}
\usepackage{amsthm}

\usepackage[capitalize,noabbrev]{cleveref}

\theoremstyle{plain}

\theoremstyle{definition}

\theoremstyle{remark}

\usepackage[textsize=tiny]{todonotes}

\icmltitlerunning{Adversarial Prompting for Robust Astronomy Hypothesis Generation}

\begin{document}

\twocolumn[
\icmltitle{Harnessing the Power of Adversarial Prompting and Large Language Models for Robust Hypothesis Generation in Astronomy}



\icmlsetsymbol{equal}{*}

\begin{icmlauthorlist}
\icmlauthor{Ioana Ciuc\u{a}}{equal,rsaa,soco}
\icmlauthor{Yuan-Sen Ting}{equal,rsaa,soco}
\icmlauthor{Sandor Kruk}{esac}
\icmlauthor{Kartheik Iyer}{col}
\end{icmlauthorlist}

\icmlaffiliation{rsaa}{Research School of Astronomy \& Astrophysics, Australian National University, Cotter Rd., Weston, ACT 2611, Australia}
\icmlaffiliation{soco}{School of Computing, Australian National University, Acton, ACT 2601, Australia}
\icmlaffiliation{esac}{European Space Astronomy Centre, European Space Agency, Villafranca del Castillo, Madrid 28692, Spain}
\icmlaffiliation{col}{Columbia Astrophysics Laboratory, Columbia University, New York, NY 10027, USA}

\icmlcorrespondingauthor{Ioana Ciuca}{Ioana.ciuca@anu.edu.au}
\icmlcorrespondingauthor{Yuan-Sen Ting}{yuan-sen.ting@anu.edu.au}

\icmlkeywords{Machine Learning, ICML}

\vskip 0.3in
]


\printAffiliationsAndNotice{\icmlEqualContribution} 

\begin{abstract}
This study investigates the application of Large Language Models (LLMs), specifically GPT-4, within Astronomy. We employ in-context prompting, supplying the model with up to 1000 papers from the NASA Astrophysics Data System, to explore the extent to which performance can be improved by immersing the model in domain-specific literature. Our findings point towards a substantial boost in hypothesis generation when using in-context prompting, a benefit that is further accentuated by adversarial prompting. We illustrate how adversarial prompting empowers GPT-4 to extract essential details from a vast knowledge base to produce meaningful hypotheses, signaling an innovative step towards employing LLMs for scientific research in Astronomy.
\end{abstract}

\section{Introduction}\label{submission}

Significant strides in Natural Language Processing (NLP) have been made possible through attention mechanisms and transformer architecture, leading to the development of Large Language Models (LLMs) such as GPT-4 \citep{Vig2019,brown2020language,Ouyang2022}. These models exhibit extraordinary aptitude in understanding, generating, and interacting with human language. They go beyond deciphering complex linguistic patterns to making non-trivial deductions and forming relationships across diverse contexts \citep[e.g.,][]{Devlin2018,Elkins2020}.

Two intriguing facets of these models have stirred excitement for their potential that surpasses their initial intended applications. Firstly, despite LLMs' propensity to sample posterior means of languages—a factor that can occasionally result in non-trivial hallucination problems—improved performance has been witnessed through in-context prompting \citep{Wang2022,Wei2022,Zhang2022}. This enhancement enables them to handle complex, domain-specific tasks \citep[e.g.,][]{Radford2018, brown2020language,Lu2022}. Secondly, these models, when combined with revolutionary technologies like {\tt Langchain}\footnote{https://python.langchain.com} to provide extensive context to the LLMs, expand their functionality across a wide range of fields.

While methods like the use of adapters \citep{He2021,Karimi2021,Lora2021} can remarkably augment performance for domain-specific tasks through fine-tuning the LLMs, these approaches often prove challenging for institutions without sufficient resources. In this study, we delve into the application of low-cost in-context prompting \citep{Chen2021,Xie2021} in the realm of astronomy.

\begin{figure*}[ht!]
    \centering
    \includegraphics[width=0.8\textwidth]{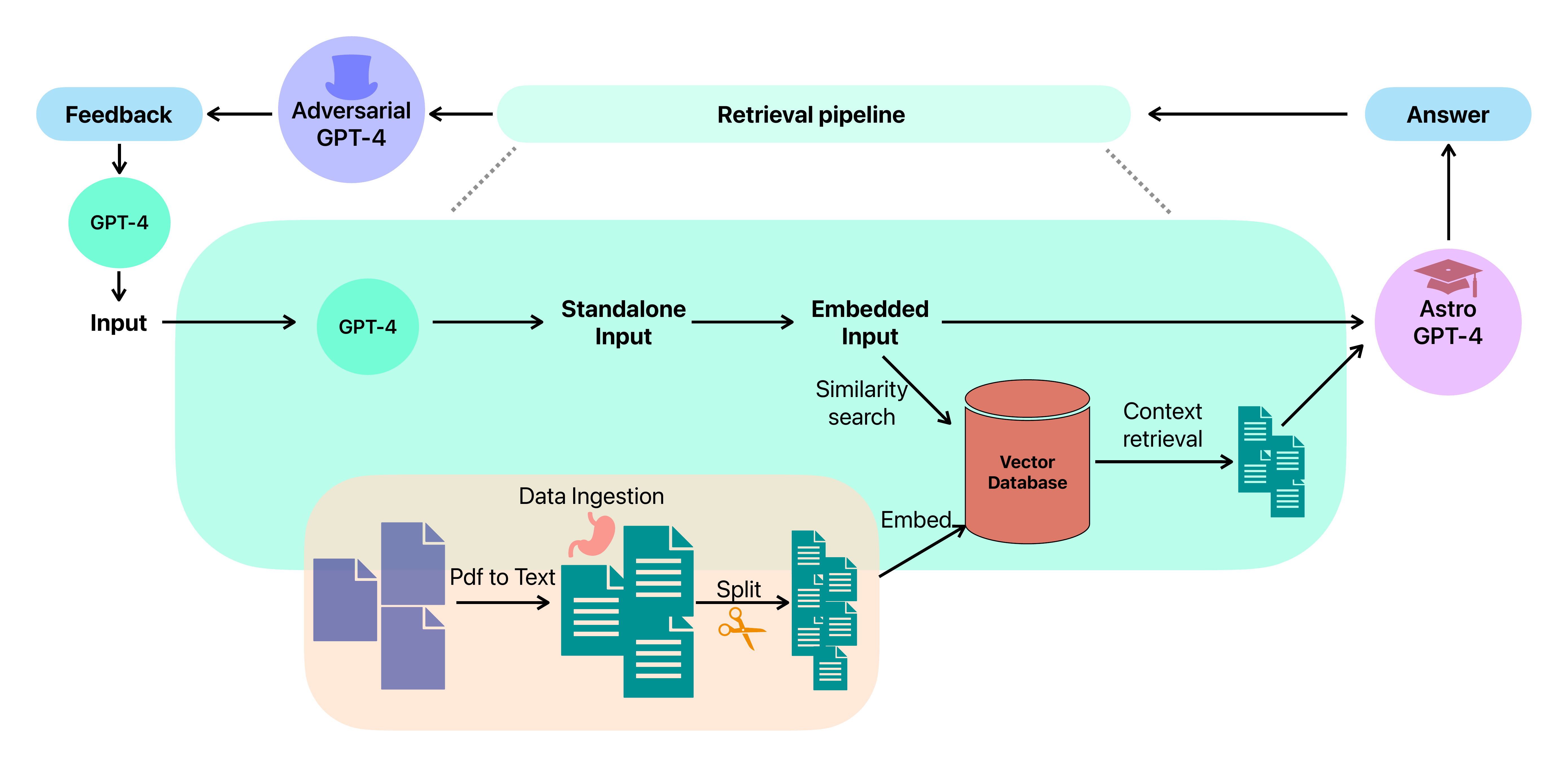}
    \vskip -0.15in
    \caption{This figure illustrates the adversarial in-context prompting workflow using OpenAI's GPT-4 model. The procedure begins with the pre-processing and embedding of Galactic Astronomy papers. A similarity search is conducted on the embedded query, and relevant document chunks are retrieved. A further contextual compression is performed to remove irrelevant information from the chunks. These compressed texts serve as input to a GPT-4 instance, which generates an idea. This idea is then critiqued by a second GPT-4 model, and the feedback is moderated by a third GPT-4 model.}
    \vskip -0.1in
    \label{fig1}
\end{figure*}

Astronomy offers a compelling case study due to three key reasons. Firstly, although the field is rich in literature, the inclusion of such text in the vast corpus used to train GPT models is probably limited. This lack leads to noticeable hallucination problems when employing naive versions of LLMs \citep{Ciuca2023}. Secondly, unlike domains that focus more on intensive, detailed studies, advancements in astronomy often stem from ``connecting the dots" across different subfields due to the universality of underlying physical processes at various scales. This feature fosters the hypothesis that extensive in-context prompting could significantly enhance hypothesis generation if LLMs are initially exposed to a broad range of literature.

Lastly, astronomy's longstanding ``open sky" policy makes it an ideal candidate for in-context prompting research. This policy ensures that most data sets are publicly available immediately or after a short proprietary period \citep{Almeida2023,Fabricius2021}. Further, the field possesses a comprehensive, well-curated literature database. The internet has enabled the archiving of astronomical knowledge, with NASA's Astrophysics Data System hosting over 15 million resources, effectively covering the entire spectrum of astronomical literature utilized by researchers \citep{Accomazzi2015,Borgman2021}. This accessibility greatly aids our engagement with the astronomy database.

\section{Literature retrieval and pre-processing}
\label{sec:data}

For this study, we focused our exploration on Galactic Astronomy, utilizing our domain expertise to assess the results. We selected Galactic Astronomy as our focal area due to its integrative nature, fusing knowledge from diverse subfields. The study of galaxy evolution not only incorporates the fundamental understanding of stars and stellar populations \citep{Aouad2020,Sanchez2022} but it is also influenced by large-scale cosmological environmental factors \citep{Singh2020,Whitney2021}. Therefore, studying galaxy evolution provides both exciting challenges and abundant possibilities for harnessing implicit knowledge embedded within the vast network of literature.

Our study includes a selection of 1,000 papers related to Galactic Astronomy from the NASA ADS \citep{Accomazzi2015} Astronomy collection. Our chosen papers were identified through a database query based on criteria such as `Gaia' appearing in the abstract, publications from the last ten years (since the Gaia launch date), being refereed journal articles, and the inclusion of relevant keywords such as `galaxy kinematics and dynamics', `galaxy structure', `galaxy disk', `galaxy halo', `galaxy abundances', and `galaxy evolution'. Our initial query yielded more than 1,000 papers, leading us to prioritize the most recent publications. Our curated collection contains details such as the ArxivID, Publication Date, Authors, Title, Abstract, Citation, and Key, providing a comprehensive dataset for our analysis. The full dataset as well as the codebase used in our analysis can be found here for reproducibility\footnote{https://github.com/errai34/IdeaGPT}.

\begin{figure*}[t]
    \centering
    \includegraphics[width=0.85\textwidth]{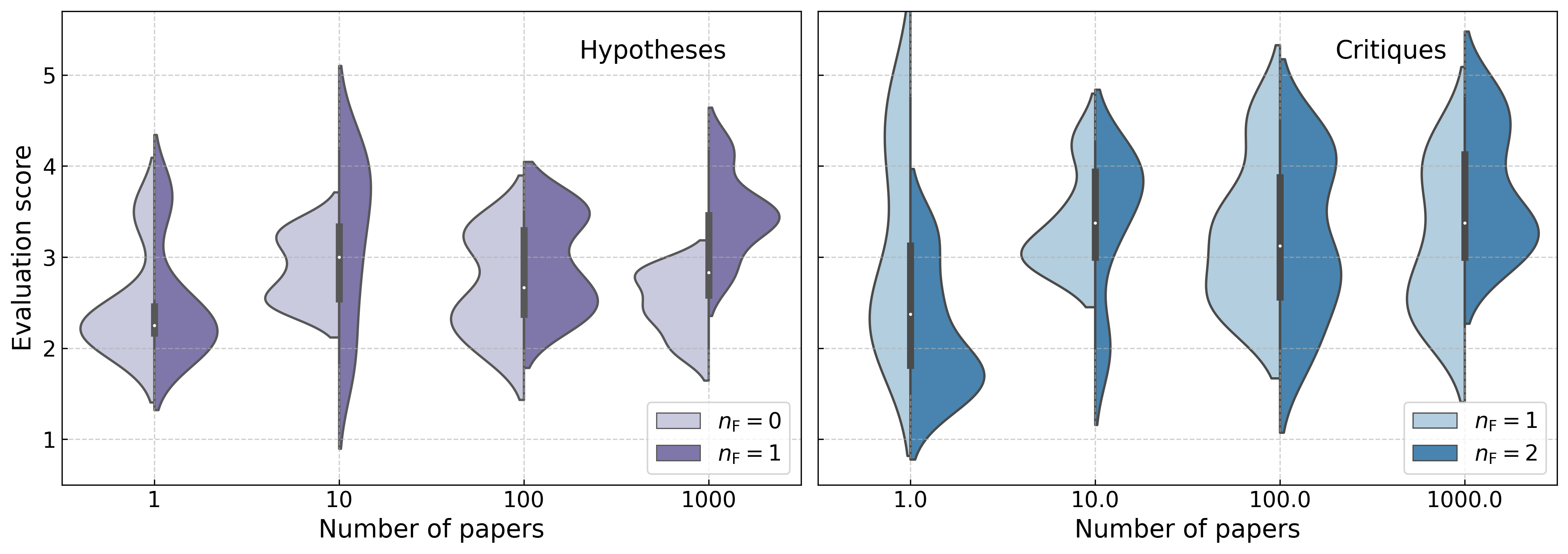}
    \vskip -0.1in
    \caption{Adversarial prompting and domain-specific context enrichment significantly enhance hypothesis generation quality. 60 hypotheses and 40 critiques generated by the AI were evaluated by two human experts, with the mean scores reported for individual instances. The iterations of adversarial prompting ($n_F$) was instrumental in driving substantial enhancements in both the quality and consistency of the AI judge and AI generator outputs, particularly when they were supplied with an extensive context ($N=1000$ papers). Crucially, in the absence of adversarial prompting ($n_F=0$), the quality of the hypothesis remained stagnant despite the provision of ample context. This observation underscores the stark contrast and superior effectiveness of adversarial prompting.}
    \vskip -0.1in
    \label{fig2}
\end{figure*}

\section{Astro-GPT Workflow}

Our exploration capitalizes on the abilities of OpenAI's GPT-4 model \citep{OpenAIGPT4}. The first step in in-context prompting involves pre-processing 1,000 papers from the Galactic Astronomy corpus using the \texttt{langchain} library. Each paper, transformed from PDF to text, is subsequently segmented into `chunks' of 1,000 tokens each. These segmented units are then embedded using OpenAI's \texttt{text-ada-002} embedding model.

The retrieval phase begins with converting the chat history and input query into a standalone input, which is then embedded. A similarity search is conducted between the embedded query and the vector database. We then use \texttt{langchain}'s contextual compression to filter out irrelevant information from the individual chunks. These final texts, combined with the standalone input, form the foundation upon which a GPT-4 model, having a context window of approximately 8,000 tokens, formulates ideas. To scrutinize the model's prowess, we design an adversarial experiment. This involves a secondary GPT-4 model that critiques the idea, highlighting its frailties and suggesting potential enhancements. This feedback is reformulated within a feedback-question structure by a third GPT-4 instance and returned to the initial model.

Implementing our experimental setup, we use $N_{k}$ papers, where $k \in \{1, 10, 100, 1000\}$. Each sample undergoes hypothesis generation by the `Generation GPT-4' instance (our in-context prompted model on $k$ papers). An adversarial response from `Adversarial GPT-4' follows, which is reformulated by a moderator GPT-4 instance and fed back to the generator model. This cycle, yielding three hypotheses and two critiques per experiment, is repeated twice for each $N_k$ and replicated five times in total. The same approach is applied to 1,000 papers, without resampling, accumulating a total of 60 hypotheses and 40 critiques.

\begin{figure*}[t]
    \centering
     \includegraphics[width=0.8\textwidth]{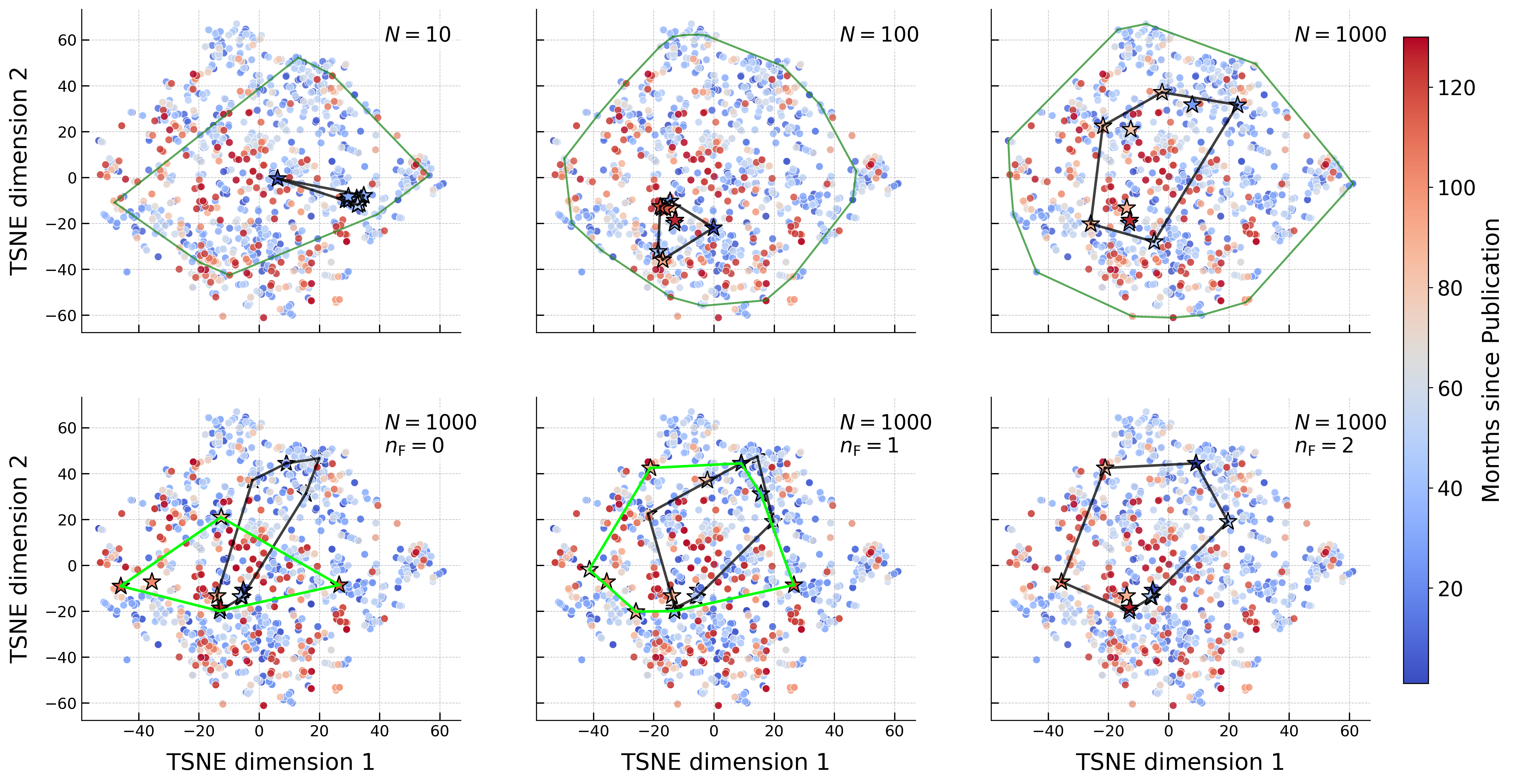}
     \vskip -0.1in
     \caption{Visual representation of each hypothesis's `knowledge footprint', depicted as black polygons within the TSNE projection of the abstracts of our corpus comprising 1000 papers. As the quantity of papers consumed increases, the model leverages a more diverse array of topics, thus boosting the quality of the hypothesis (as seen in the top panels). The green hull in the top panel shows the overall knowledge base that the model have access through in-context prompting. The lower panel demonstrates how the `knowledge footprint' evolves with varying quantities $n_F$ of adversarial attacks in the case of 1000 papers. The black polygon signifies the footprint of the original response, while the lime indicates the information utilized by the AI judge for critique.}
     \vskip -0.1in
     \label{fig3}
\end{figure*}

\section{Results}

\subsection{Human Evaluation}

Given the qualitative nature of hypothesis generation, we needed an assessment process that, while inherently subjective, would match the expectations of human experts. For this, we involved two domain experts in the field of Galactic Astronomy to evaluate the quality of the generated hypotheses. These were graded based on the number of papers included within the domain-specific context, and we computed the average score from these dual-human evaluations for each hypothesis. The hypotheses are graded based on a rubric of three categories -- scientific accuracy, creativity and feasibility, and the average score of these three domains assumed to be the final score. We also evaluated the critiques provided by the AI judge, which had access to the same contextual information.

As illustrated in the left panel of Fig.~\ref{fig2}, adversarial prompting proved to be a critical tool in markedly improving hypothesis generation. The quality of hypothesis generation, without adversarial prompting, showed little dependence on the number of papers, suggesting that in-context prompting alone, while helpful for mitigating hallucination, did not suffice for a comprehensive understanding of the corpus. 

The introduction of adversarial prompting considerably altered this outcome. A significant improvement in the quality of hypothesis generation was observed both for the AI generator and the AI judge, even without explicitly aligning the models with human expectations. Notably, adversarial prompting introduced a strong correlation between hypothesis quality and the number of papers reviewed, especially at larger context ($N=1000$). It also leads to a much more consistent in term of the quality of the hypotheses (and the critiques). The average quality score rose significantly from 2.5 (when 10 papers were used as context, where a score of 3/5 corresponds to a typical hypothesis by a competent PhD student) to a near-expert level of 4/5 when 1,000 papers were included, emphasizing the potential of adversarial prompting in enhancing the quality of scientific hypothesis generation. We refer to the Appendix for examples.

\subsection{Exploration of Embeddings}

To truly understand the power of adversarial prompting, we first passed the abstracts of our set of 1000 astronomy papers through the \texttt{text-ada-002} embedding model and arranged them into a 2D TSNE projection. This captured the contextual differences and similarities of these 1000 papers.

For each hypothesis generated, we determined which papers inspired it by querying the GPT model. In Fig.~\ref{fig3}, we visualized this `knowledge footprint' for each hypothesis as black polygons within a green hull representing all the papers GPT-4 had access to. From Fig.~\ref{fig3}'s top panels, it's clear that as the number of papers grows, GPT broadens its scope, drawing on diverse topics to build interconnected hypotheses. We note that, even with a small training pool of 10 papers, we design our experiment in a way that the corpus still covers a wide range of topics, but GPT-4 lacks the context to connect them in a meaningful way, leading to more general hypotheses (see Appendix).

The bottom panel in Fig.~\ref{fig3}, concentrating on the case with 1000 papers, explores how the knowledge footprint evolves with different numbers of adversarial attacks. In the preliminary iteration (bottom left), the judge ingeniously identifies areas of critique based on knowledge overlooked by the original response. This compels the generator to expand its scope further (as shown in the bottom middle and right panels), to appropriately address the criticism. As shown in some of the examples in the Appendix, adversarial prompting allows the GPT-4 model to genuinely benefit from a large number of contextual inputs, guiding the model towards a more coherent understanding of the topic, rather than creating a mere `fact jumble'. The specific examples of hypotheses and corresponding critiques are shown in the Appendix.

\section{Conclusion and Future Direction}

In this research, we delved into a detailed examination of the GPT-4 model's ability to propose novel hypotheses in the domain of Galactic Astronomy using in-context prompting. Our findings confirm that in-context prompting significantly mitigates hallucination, leading to the generation of meaningful hypotheses that can compete with substantive thesis topics, as evaluated by domain experts. Importantly, we found that the inclusion of adversarial prompts enables the model to progressively enhance its performance based on the number of contextual papers. However, a naive implementation of in-context prompting without adversarial judgment fails to replicate this improvement.

While our study marks the inception of a pioneering exploration of in-context prompting for scientific hypothesis generation, it's clear that this dynamic field is rapidly evolving. Thus, we have identified several crucial areas for enhancement. These areas include (a) an improved and automated evaluation method for hypotheses. We have observed that while the AI judge can assist the AI generator, improvements are primarily in technical detail rather than deep insights. We propose leveraging well-curated question-and-answer pairs \citep[e.g.][]{Dugan2022} to better align the judge with human expectations. (b) Instead of focusing solely on hypothesis generation, integrating other downstream tasks and finite fine-tuning models with smaller adapter models could potentially improve inferences. We have commenced curating metadata from ADS to better design these tasks.

\section{Broader Impact}

In this study, our focus on in-context prompting, rather than the more computationally intensive fine-tuning, is inspired by the aim to democratize the utilization of LLMs for scientific inquiry. Current GPT models, due to their immense parameter sets, often render fine-tuning impractical. In the era of Large Language Models, it is crucial to determine whether all academic institutions, regardless of size or available computational resources, can keep pace with these rapidly advancing technologies.

This critical inquiry forms the crux of our study, and our findings present a hopeful outlook. Our study indicates that, with the right strategies and approaches with `humans in the loop' as domain experts, the barrier to harnessing the full power of these advanced LLMs can be lowered. As a result, we envision a future where all institutions, regardless of size or resources, can contribute to and benefit from the swift advancements in AI, enhancing the collective endeavour of scientific discovery. Our journey into this new frontier of Large Language Models is just beginning, and it promises a thrilling ride full of unexpected insights and revolutionary breakthroughs.

\bibliography{ms}

\begin{thebibliography}{26}
\providecommand{\natexlab}[1]{#1}
\providecommand{\url}[1]{\texttt{#1}}
\expandafter\ifx\csname urlstyle\endcsname\relax
  \providecommand{\doi}[1]{doi: #1}\else
  \providecommand{\doi}{doi: \begingroup \urlstyle{rm}\Url}\fi

\bibitem[{Accomazzi} et~al.(2015){Accomazzi}, {Kurtz}, {Henneken}, {Chyla},
  {Luker}, {Grant}, {Thompson}, {Holachek}, {Dave}, and
  {Murray}]{Accomazzi2015}
{Accomazzi}, A., {Kurtz}, M.~J., {Henneken}, E.~A., {Chyla}, R., {Luker}, J.,
  {Grant}, C.~S., {Thompson}, D.~M., {Holachek}, A., {Dave}, R., and {Murray},
  S.~S.
\newblock {ADS: The Next Generation Search Platform}.
\newblock In {Holl}, A., {Lesteven}, S., {Dietrich}, D., and {Gasperini}, A.
  (eds.), \emph{Open Science at the Frontiers of Librarianship}, volume 492 of
  \emph{Astronomical Society of the Pacific Conference Series}, pp.\  189,
  April 2015.
\newblock \doi{10.48550/arXiv.1503.04194}.

\bibitem[Almeida et~al.(2023)Almeida, Anderson, Argudo-Fernández, Badenes,
  Barger, Barrera-Ballesteros, Bender, Benitez, Besser, Bizyaev, Blanton,
  Bochanski, Bovy, Brandt, Brownstein, Buchner, Bulbul, Burchett, Díaz,
  Carlberg, Casey, Chandra, Cherinka, Chiappini, Coker, Comparat, Conroy,
  Contardo, Cortes, Covey, Crane, Cunha, Dabbieri, au2, Davis, Lee, Delgado,
  Demasi, Mille, Donor, Dow, Dwelly, Eracleous, Eriksen, Fan, Farr, Frederick,
  Fries, Frinchaboy, Gaensicke, Ge, Ávila, Grabowski, Grier, Guiglion, Gupta,
  Hall, Hawkins, Hayes, Hermes, Hernández-García, Hogg, Holtzman,
  Ibarra-Medel, Ji, Jofre, Johnson, Jones, Kinemuchi, Kluge, Koekemoer,
  Kollmeier, Kounkel, Krishnarao, Krumpe, Lacerna, Lago, Laporte, Liu, Liu,
  Liu, Lopes, Macktoobian, Malanushenko, Maoz, Masseron, Masters, Matijevic,
  McBride, Medan, Merloni, Morrison, Myers, Mészáros, Negrete, Nidever,
  Nitschelm, Oravetz, Oravetz, Pan, Peng, Pinsonneault, Pogge, Qiu,
  de~Andrade~Queiroz, Ramirez, Rix, Rosso, Runnoe, Salvato, Sanchez, Santana,
  Saydjari, Sayres, Schlaufman, Schneider, Schwope, Serna, Shen, Sobeck, Song,
  Souto, Spoo, Stassun, Steinmetz, Straumit, Stringfellow, Sánchez-Gallego,
  Taghizadeh-Popp, Tayar, Thakar, Tissera, Tkachenko, Toledo, Trakhtenbrot,
  Trincado, Troup, Trump, Tuttle, Ulloa, Vazquez-Mata, Alfaro, Villanova,
  Wachter, Weijmans, Wheeler, Wilson, Wojno, Wolf, Xue, Ybarra, Zari, and
  Zasowski]{Almeida2023}
Almeida, A., Anderson, S.~F., Argudo-Fernández, M., Badenes, C., Barger, K.,
  Barrera-Ballesteros, J.~K., Bender, C.~F., Benitez, E., Besser, F., Bizyaev,
  D., Blanton, M.~R., Bochanski, J., Bovy, J., Brandt, W.~N., Brownstein,
  J.~R., Buchner, J., Bulbul, E., Burchett, J.~N., Díaz, M.~C., Carlberg,
  J.~K., Casey, A.~R., Chandra, V., Cherinka, B., Chiappini, C., Coker, A.~A.,
  Comparat, J., Conroy, C., Contardo, G., Cortes, A., Covey, K., Crane, J.~D.,
  Cunha, K., Dabbieri, C., au2, J. W. D.~J., Davis, M.~C., Lee, N.~D., Delgado,
  J. E.~M., Demasi, S., Mille, F.~D., Donor, J., Dow, P., Dwelly, T.,
  Eracleous, M., Eriksen, J., Fan, X., Farr, E., Frederick, S., Fries, L.,
  Frinchaboy, P., Gaensicke, B.~T., Ge, J., Ávila, C.~G., Grabowski, K.,
  Grier, C., Guiglion, G., Gupta, P., Hall, P., Hawkins, K., Hayes, C.~R.,
  Hermes, J.~J., Hernández-García, L., Hogg, D.~W., Holtzman, J.~A.,
  Ibarra-Medel, H.~J., Ji, A., Jofre, P., Johnson, J.~A., Jones, A.~M.,
  Kinemuchi, K., Kluge, M., Koekemoer, A., Kollmeier, J.~A., Kounkel, M.,
  Krishnarao, D., Krumpe, M., Lacerna, I., Lago, P. J.~A., Laporte, C., Liu,
  A., Liu, C., Liu, X., Lopes, A.~R., Macktoobian, M., Malanushenko, V., Maoz,
  D., Masseron, T., Masters, K.~L., Matijevic, G., McBride, A., Medan, I.,
  Merloni, A., Morrison, S., Myers, N., Mészáros, S., Negrete, C.~A.,
  Nidever, D.~L., Nitschelm, C., Oravetz, A., Oravetz, D., Pan, K., Peng, Y.,
  Pinsonneault, M.~H., Pogge, R., Qiu, D., de~Andrade~Queiroz, A.~B., Ramirez,
  S.~V., Rix, H.-W., Rosso, D.~F., Runnoe, J., Salvato, M., Sanchez, S.~F.,
  Santana, F.~A., Saydjari, A., Sayres, C., Schlaufman, K.~C., Schneider,
  D.~P., Schwope, A., Serna, J., Shen, Y., Sobeck, J., Song, Y.-Y., Souto, D.,
  Spoo, T., Stassun, K.~G., Steinmetz, M., Straumit, I., Stringfellow, G.,
  Sánchez-Gallego, J., Taghizadeh-Popp, M., Tayar, J., Thakar, A., Tissera,
  P.~B., Tkachenko, A., Toledo, H.~H., Trakhtenbrot, B., Trincado, J. G.~F.,
  Troup, N., Trump, J.~R., Tuttle, S., Ulloa, N., Vazquez-Mata, J.~A., Alfaro,
  P.~V., Villanova, S., Wachter, S., Weijmans, A.-M., Wheeler, A., Wilson, J.,
  Wojno, L., Wolf, J., Xue, X.-X., Ybarra, J.~E., Zari, E., and Zasowski, G.
\newblock The eighteenth data release of the sloan digital sky surveys:
  Targeting and first spectra from sdss-v, 2023.

\bibitem[Aouad et~al.(2020)Aouad, James, and Chilingarian]{Aouad2020}
Aouad, C.~J., James, P.~A., and Chilingarian, I.~V.
\newblock Coupling local to global star formation in spiral galaxies: the
  effect of differential rotation.
\newblock \emph{Monthly Notices of the Royal Astronomical Society},
  496\penalty0 (4):\penalty0 5211--5226, jul 2020.
\newblock \doi{10.1093/mnras/staa1945}.

\bibitem[{Borgman} \& {Wofford}(2021){Borgman} and {Wofford}]{Borgman2021}
{Borgman}, C.~L. and {Wofford}, M.~F.
\newblock {From Data Processes to Data Products: Knowledge Infrastructures in
  Astronomy}.
\newblock \emph{arXiv e-prints}, art. arXiv:2109.01707, September 2021.
\newblock \doi{10.48550/arXiv.2109.01707}.

\bibitem[Brown et~al.(2020)Brown, Mann, Ryder, Subbiah, Kaplan, Dhariwal,
  Neelakantan, Shyam, Sastry, Askell, Agarwal, Herbert-Voss, Krueger, Henighan,
  Child, Ramesh, Ziegler, Wu, Winter, Hesse, Chen, Sigler, Litwin, Gray, Chess,
  Clark, Berner, McCandlish, Radford, Sutskever, and Amodei]{brown2020language}
Brown, T.~B., Mann, B., Ryder, N., Subbiah, M., Kaplan, J., Dhariwal, P.,
  Neelakantan, A., Shyam, P., Sastry, G., Askell, A., Agarwal, S.,
  Herbert-Voss, A., Krueger, G., Henighan, T., Child, R., Ramesh, A., Ziegler,
  D.~M., Wu, J., Winter, C., Hesse, C., Chen, M., Sigler, E., Litwin, M., Gray,
  S., Chess, B., Clark, J., Berner, C., McCandlish, S., Radford, A., Sutskever,
  I., and Amodei, D.
\newblock Language models are few-shot learners, 2020.

\bibitem[Chen et~al.(2021)Chen, Zhong, Zha, Karypis, and He]{Chen2021}
Chen, Y., Zhong, R., Zha, S., Karypis, G., and He, H.
\newblock Meta-learning via language model in-context tuning.
\newblock \emph{arXiv preprint arXiv:2110.07814}, 2021.

\bibitem[{Ciuc{\u{a}}} et~al.(2023){Ciuc{\u{a}}}, {Kawata}, {Ting}, {Grand},
  {Miglio}, {Hayden}, {Baba}, {Fragkoudi}, {Monty}, {Buder}, and
  {Freeman}]{Ciuca2023}
{Ciuc{\u{a}}}, I., {Kawata}, D., {Ting}, Y.-S., {Grand}, R. J.~J., {Miglio},
  A., {Hayden}, M., {Baba}, J., {Fragkoudi}, F., {Monty}, S., {Buder}, S., and
  {Freeman}, K.
\newblock {Chasing the impact of the Gaia-Sausage-Enceladus merger on the
  formation of the Milky Way thick disc}.
\newblock \emph{\mnras}, March 2023.
\newblock \doi{10.1093/mnrasl/slad033}.

\bibitem[{Devlin} et~al.(2018){Devlin}, {Chang}, {Lee}, and
  {Toutanova}]{Devlin2018}
{Devlin}, J., {Chang}, M.-W., {Lee}, K., and {Toutanova}, K.
\newblock {BERT: Pre-training of Deep Bidirectional Transformers for Language
  Understanding}.
\newblock \emph{arXiv e-prints}, art. arXiv:1810.04805, October 2018.
\newblock \doi{10.48550/arXiv.1810.04805}.

\bibitem[Dugan et~al.(2022)Dugan, Miltsakaki, Upadhyay, Ginsberg, Gonzalez,
  Choi, Yuan, and Callison-Burch]{Dugan2022}
Dugan, L., Miltsakaki, E., Upadhyay, S., Ginsberg, E., Gonzalez, H., Choi, D.,
  Yuan, C., and Callison-Burch, C.
\newblock A feasibility study of answer-agnostic question generation for
  education.
\newblock In \emph{Findings of the Association for Computational Linguistics:
  ACL 2022}, pp.\  1919--1926, Dublin, Ireland, May 2022. Association for
  Computational Linguistics.
\newblock \doi{10.18653/v1/2022.findings-acl.151}.

\bibitem[Elkins \& Chun(2020)Elkins and Chun]{Elkins2020}
Elkins, K. and Chun, J.
\newblock Can gpt-3 pass a writer’s turing test?
\newblock \emph{Journal of Cultural Analytics}, 5\penalty0 (2), 2020.

\bibitem[Fabricius et~al.(2021)Fabricius, Luri, Arenou, Babusiaux, Helmi,
  Muraveva, Reyl{\'{e} }, Spoto, Vallenari, Antoja, Balbinot, Barache, Bauchet,
  Bragaglia, Busonero, Cantat-Gaudin, Carrasco, Diakit{\'{e}}, Fabrizio,
  Figueras, Garcia-Gutierrez, Garofalo, Jordi, Kervella, Khanna, Leclerc,
  Licata, Lambert, Marrese, Masip, Ramos, Robichon, Robin, Romero-G{\'{o}}mez,
  Rubele, and Weiler]{Fabricius2021}
Fabricius, C., Luri, X., Arenou, F., Babusiaux, C., Helmi, A., Muraveva, T.,
  Reyl{\'{e} }, C., Spoto, F., Vallenari, A., Antoja, T., Balbinot, E.,
  Barache, C., Bauchet, N., Bragaglia, A., Busonero, D., Cantat-Gaudin, T.,
  Carrasco, J.~M., Diakit{\'{e}}, S., Fabrizio, M., Figueras, F.,
  Garcia-Gutierrez, A., Garofalo, A., Jordi, C., Kervella, P., Khanna, S.,
  Leclerc, N., Licata, E., Lambert, S., Marrese, P.~M., Masip, A., Ramos, P.,
  Robichon, N., Robin, A.~C., Romero-G{\'{o}}mez, M., Rubele, S., and Weiler,
  M.
\newblock $\less$i$\greater$gaia$\less$/i$\greater$early data release 3.
\newblock \emph{Astronomy {\&}amp$\mathsemicolon$ Astrophysics}, 649:\penalty0
  A5, apr 2021.
\newblock \doi{10.1051/0004-6361/202039834}.

\bibitem[He et~al.(2021)He, Liu, Ye, Tan, Ding, Cheng, Low, Bing, and
  Si]{He2021}
He, R., Liu, L., Ye, H., Tan, Q., Ding, B., Cheng, L., Low, J.-W., Bing, L.,
  and Si, L.
\newblock On the effectiveness of adapter-based tuning for pretrained language
  model adaptation.
\newblock \emph{arXiv preprint arXiv:2106.03164}, 2021.

\bibitem[{Hu} et~al.(2021){Hu}, {Shen}, {Wallis}, {Allen-Zhu}, {Li}, {Wang},
  {Wang}, and {Chen}]{Lora2021}
{Hu}, E.~J., {Shen}, Y., {Wallis}, P., {Allen-Zhu}, Z., {Li}, Y., {Wang}, S.,
  {Wang}, L., and {Chen}, W.
\newblock {LoRA: Low-Rank Adaptation of Large Language Models}.
\newblock \emph{arXiv e-prints}, art. arXiv:2106.09685, June 2021.
\newblock \doi{10.48550/arXiv.2106.09685}.

\bibitem[Karimi~Mahabadi et~al.(2021)Karimi~Mahabadi, Henderson, and
  Ruder]{Karimi2021}
Karimi~Mahabadi, R., Henderson, J., and Ruder, S.
\newblock Compacter: Efficient low-rank hypercomplex adapter layers.
\newblock \emph{Advances in Neural Information Processing Systems},
  34:\penalty0 1022--1035, 2021.

\bibitem[Lu et~al.(2022)Lu, Mishra, Xia, Qiu, Chang, Zhu, Tafjord, Clark, and
  Kalyan]{Lu2022}
Lu, P., Mishra, S., Xia, T., Qiu, L., Chang, K.-W., Zhu, S.-C., Tafjord, O.,
  Clark, P., and Kalyan, A.
\newblock Learn to explain: Multimodal reasoning via thought chains for science
  question answering.
\newblock In Koyejo, S., Mohamed, S., Agarwal, A., Belgrave, D., Cho, K., and
  Oh, A. (eds.), \emph{Advances in Neural Information Processing Systems},
  volume~35, pp.\  2507--2521. Curran Associates, Inc., 2022.

\bibitem[OpenAI(2023)]{OpenAIGPT4}
OpenAI.
\newblock Gpt-4 technical report.
\newblock \emph{ArXiv}, abs/2303.08774, 2023.

\bibitem[{Ouyang} et~al.(2022){Ouyang}, {Wu}, {Jiang}, {Almeida}, {Wainwright},
  {Mishkin}, {Zhang}, {Agarwal}, {Slama}, {Ray}, {Schulman}, {Hilton},
  {Kelton}, {Miller}, {Simens}, {Askell}, {Welinder}, {Christiano}, {Leike},
  and {Lowe}]{Ouyang2022}
{Ouyang}, L., {Wu}, J., {Jiang}, X., {Almeida}, D., {Wainwright}, C.~L.,
  {Mishkin}, P., {Zhang}, C., {Agarwal}, S., {Slama}, K., {Ray}, A.,
  {Schulman}, J., {Hilton}, J., {Kelton}, F., {Miller}, L., {Simens}, M.,
  {Askell}, A., {Welinder}, P., {Christiano}, P., {Leike}, J., and {Lowe}, R.
\newblock {Training language models to follow instructions with human
  feedback}.
\newblock \emph{arXiv e-prints}, art. arXiv:2203.02155, March 2022.
\newblock \doi{10.48550/arXiv.2203.02155}.

\bibitem[Radford \& Narasimhan(2018)Radford and Narasimhan]{Radford2018}
Radford, A. and Narasimhan, K.
\newblock Improving language understanding by generative pre-training.
\newblock 2018.

\bibitem[{S{\'a}nchez} et~al.(2022){S{\'a}nchez}, {G{\'o}mez Medina},
  {Barrera-Ballesteros}, {Galbany}, {Bolatto}, and {Wong}]{Sanchez2022}
{S{\'a}nchez}, S.~F., {G{\'o}mez Medina}, D.~C., {Barrera-Ballesteros}, J.~K.,
  {Galbany}, L., {Bolatto}, A., and {Wong}, T.
\newblock {The local and global relations between $\Sigma_\star$ , $\Sigma_{\rm
  SFR}$ and $\Sigma_{\rm mol}$ that regulate star-formation}.
\newblock \emph{arXiv e-prints}, art. arXiv:2212.03738, December 2022.
\newblock \doi{10.48550/arXiv.2212.03738}.

\bibitem[Singh et~al.(2020)Singh, Mahajan, and Bagla]{Singh2020}
Singh, A., Mahajan, S., and Bagla, J.~S.
\newblock Study of galaxies on large-scale filaments in simulations.
\newblock \emph{Monthly Notices of the Royal Astronomical Society},
  497\penalty0 (2):\penalty0 2265--2275, jul 2020.
\newblock \doi{10.1093/mnras/staa1913}.

\bibitem[Vig(2019)]{Vig2019}
Vig, J.
\newblock Visualizing attention in transformer-based language representation
  models.
\newblock \emph{arXiv preprint arXiv:1904.02679}, 2019.

\bibitem[Wang et~al.(2022)Wang, Wei, Schuurmans, Le, Chi, and Zhou]{Wang2022}
Wang, X., Wei, J., Schuurmans, D., Le, Q., Chi, E., and Zhou, D.
\newblock Self-consistency improves chain of thought reasoning in language
  models.
\newblock \emph{arXiv preprint arXiv:2203.11171}, 2022.

\bibitem[Wei et~al.(2022)Wei, Wang, Schuurmans, Bosma, Chi, Le, and
  Zhou]{Wei2022}
Wei, J., Wang, X., Schuurmans, D., Bosma, M., Chi, E., Le, Q., and Zhou, D.
\newblock Chain of thought prompting elicits reasoning in large language
  models.
\newblock \emph{arXiv preprint arXiv:2201.11903}, 2022.

\bibitem[Whitney et~al.(2021)Whitney, Ferreira, Conselice, and
  Duncan]{Whitney2021}
Whitney, A., Ferreira, L., Conselice, C.~J., and Duncan, K.
\newblock Galaxy evolution in all five {CANDELS} fields and {IllustrisTNG}:
  Morphological, structural, and the major merger evolution to z $\sim$ 3.
\newblock \emph{The Astrophysical Journal}, 919\penalty0 (2):\penalty0 139, oct
  2021.
\newblock \doi{10.3847/1538-4357/ac1422}.

\bibitem[Xie et~al.(2021)Xie, Raghunathan, Liang, and Ma]{Xie2021}
Xie, S.~M., Raghunathan, A., Liang, P., and Ma, T.
\newblock An explanation of in-context learning as implicit bayesian inference.
\newblock \emph{arXiv preprint arXiv:2111.02080}, 2021.

\bibitem[Zhang et~al.(2022)Zhang, Zhang, Li, and Smola]{Zhang2022}
Zhang, Z., Zhang, A., Li, M., and Smola, A.
\newblock Automatic chain of thought prompting in large language models.
\newblock \emph{arXiv preprint arXiv:2210.03493}, 2022.

\end{thebibliography}
\bibliographystyle{icml2023}

\newpage
\appendix
\onecolumn

\section{Exemplary Hypotheses and Critiques Generated}\label{app:examples}

This section presents three types of hypotheses generated through our in-context prompting and adversarial prompting techniques, along with the associated critiques provided by our adversarial GPT-4 model, which led to substantial improvements in hypothesis generation.

The first example demonstrates a hypothesis generated after subjecting the model to in-context prompting through 1000 papers, followed by two rounds of adversarial attacks.

\begin{tcolorbox}[colback=red!5!white,colframe=red!75!black]
I propose an investigation into the role of dark matter mini-halos in the formation and evolution of globular clusters (GCs) in the Milky Way. This study would combine the high-precision astrometric data from Gaia EDR3 and DR3 (Gaia Collaboration et al. 2016, 2020) with spectroscopic data from large ground-based surveys like APOGEE, GALAH, SDSS SEGUE, and LAMOST to characterize the dynamics of stars in the peripheral regions of GCs. The aim would be to determine whether these GCs are embedded in dark matter mini-halos, which could provide critical insights into their origins (Peebles 1984; Peñarrubia et al. 2017).  The proposed research would build upon the probabilistic approach developed by Kuzma et al. (2021) for studying the peripheral regions of GCs, which utilizes a mixture model in spatial and proper motion space to model cluster, extra-tidal, and contaminant stellar populations. By extending this approach to include the effects of dark matter mini-halos on the kinematics of stars in GC outskirts, we can test the hypothesis that dark matter plays a significant role in the formation and evolution of GCs.  Furthermore, this study would provide a better understanding of the distribution and properties of dark matter in the Milky Way, contributing to the broader field of near-field cosmology. Integrating this information with the existing knowledge of the hierarchical assembly of the Milky Way (Viswanathan et al. 2023) and the role of rapid gas accretion in the inner Galactic disc (Snaith et al. 2021) would help paint a more comprehensive picture of our Galaxy's formation history and its underlying dark matter distribution.
\end{tcolorbox}

This hypothesis received high approval from the human jury due to its broad contextual understanding of Galactic Astronomy, a concrete thesis proposal, appropriate references for idea origins, and building upon existing methodologies and studies.

Conversely, the following hypothesis was produced when the in-context prompting was restricted to 10 papers. Hypotheses based on limited exposure to literature often appear generic, lacking substantial insights beyond minor improvements to existing work.

\begin{tcolorbox}[colback=blue!5!white,colframe=blue!75!black]
I propose a study investigating the role of intermediate-mass gas-rich mergers in the chemical and
kinematic evolution of the Milky Way-like galaxies across a range of redshifts. This research would
build upon the findings of Ciucă et al. (2023), which suggest that the Gaia-Sausage-Enceladus (GSE)
merger significantly impacted the formation of the Galactic disc and the transition from the thick
disc to the thin disc.  The study could involve a combination of observational data from large
spectroscopic surveys such as APOGEE, GALAH, and Gaia-ESO, as well as the utilization of high-
resolution cosmological simulations like the Auriga project. Key objectives would include
characterizing the chemical abundance patterns and kinematic properties of stars in the disc and
halo components during and after the mergers and probing the impact of these mergers on the radial
distribution and migration of stars across the Galactic disc.  By focusing on intermediate-mass gas-
rich mergers, this research would fill a knowledge gap in our understanding of the role these events
play in shaping the structure and evolution of Milky Way-like galaxies. The findings would also
provide valuable insights for refining galaxy formation models and contribute to a more
comprehensive understanding of the complex history of our Galaxy.
\end{tcolorbox}

Without sufficient context, the model tends to extrapolate based on existing literature, resulting in incremental, less innovative suggestions.

Importantly, despite the exposure to a corpus of 1000 papers, without adversarial attacks, the hypotheses can often seem like a jumble of facts, as shown in the example below:

\begin{tcolorbox}[colback=green!5!white,colframe=green!75!black]
I propose a comprehensive, multi-wavelength exploration of the uncharted regions of the Galactic plane, particularly in the South Galactic cap, including areas below a declination of -65°. This astronomical research initiative can be achieved by combining existing large photometric, spectroscopic, and astrometric surveys, such as Gaia, LAMOST, APOGEE, and GALAH, with new high- resolution surveys aimed at mapping the uncovered areas of the Milky Way. 
This undertaking should prioritize the detection and characterization of previously undiscovered ultra-faint dwarf galaxies, chemically peculiar stars, and stellar streams, as well as the evaluation of their origins and potential implications for our understanding of the Milky Way's formation and evolution.
This research proposal is inspired by the unexplored regions mentioned in Antoja et al. (2015), along with the possibilities for improvement suggested by Gorski and Barmby (2020) and the advancements in large-scale surveys discussed in studies such as Yuan et al. (2014) and An and Beers (2020). Expanding our knowledge of the uncharted regions of the Galactic plane will significantly contribute to our understanding of the structure, formation, and evolution of the Milky Way. Additionally, it will allow astronomers to identify and assess previously unknown populations of stars and their kinematic properties, further refining our knowledge of the various components and phenomena that shape our Galaxy.
\end{tcolorbox}

Demonstrating adversarial prompting is critical to ensure that the GPT can harness the broad context provided in a meaningful way.

Finally, an example of high-quality critique provided by the adversarial GPT-4 model is showcased. These adversarial critiques, after being primed with 1000 papers, are highly accurate and can rival the quality of expert human critique.

\begin{tcolorbox}[colback=orange!5!white,colframe=orange!75!black]
The interdisciplinary approach proposed aims to provide a comprehensive understanding of the Milky
Way's gravitational potential by combining photometric data, high-resolution simulations, and novel
techniques to analyze phase-space spirals. However, there are several weaknesses and limitations to
consider.  First, the proposal relies heavily on photometric data, which, although less affected by
sampling biases, may still be prone to uncertainties due to dust obscuration and crowding (Rix and
Bovy 2013). Additionally, the accuracy of photometric metallicity measurements may be limited
compared to spectroscopic data (An and Beers 2021).  Second, while high-resolution simulations may
enable better modeling of the Milky Way's potential, they are computationally expensive, and current
mass and spatial resolutions may not be sufficient for accurate analysis (Darling and Widrow 2018).
Furthermore, the complexity of the Milky Way, including its non-steady-state nature and the presence
of various structures such as spiral arms and bars, adds challenges to the modeling process (Kipper
et al. 2018).
Third, novel techniques to analyze phase-space spirals may be subject to different
sources of systematic bias (Widmark et al. 2021). The proposal does not fully address how these
biases will be accounted for or the potential limitations of these techniques in the context of a
complex and evolving galaxy.
Lastly, the proposal does not extensively discuss the implications of
various selection biases in spectroscopic surveys (Sanders and Das 2018), nor the challenges in
incorporating the complexity of our galaxy in large-scale cosmological models (Everall et al.
2021c).  In conclusion, while the interdisciplinary approach has merits, further work is needed to
address these weaknesses and limitations for a more accurate and comprehensive understanding of the
Milky Way's gravitational potential.
\end{tcolorbox}

\end{document}